\documentclass[10pt,letterpaper,twocolumn]{article} 

\usepackage{ol2}
\usepackage[draft]{hyperref}
\usepackage{amsmath}
\usepackage{color}

\begin{document}

\twocolumn[ 

\title{High-efficiency frequency doubling of continuous-wave laser light}


\author{Stefan Ast,$^1$ Ramon Moghadas Nia$^1$, Axel Sch\"{o}nbeck$^1$, Nico Lastzka,$^1$, Jessica Steinlechner,$^1$\linebreak Tobias Eberle,$^{1,2}$ Moritz Mehmet,$^{1,2}$ Sebastian Steinlechner$^1$ and Roman Schnabel$^{1,*}$}

\address{
$^1$Max-Planck-Institute for Gravitational Physics, Albert-Einstein-Institute, and Institut f\"{u}r Gravitationsphysik,
Leibniz Universit\"{a}t Hannover, Callinstrasse 38, D-30167 Hannover, Germany
\\
$^2$Centre for Quantum Engineering and Space-Time Research – QUEST, Leibniz Universit\"{a}t Hannover, Welfengarten 1, 30167 Hannover, Germany
\\
$^*$Corresponding author: roman.schnabel@aei.mpg.de
}

\begin{abstract}
We report on the observation of high efficiency frequency doubling of 1550\,nm continuous-wave laser light in a nonlinear cavity containing a periodically poled potassium titanyl phosphate crystal (\mbox{PPKTP}). The fundamental field had a power of $1.10\,\text{W}$ and was converted into $1.05\,\text{W}$ at 775\,nm, yielding a total external conversion efficiency of $95\pm1\,\%$. The latter value is based on the measured depletion of the fundamental field being consistent with the absolute values derived from numerical simulations. 
According to our model, the conversion efficiency achieved was limited by the non-perfect mode-matching into the nonlinear cavity and by the non-perfect impedance matching for the maximum input power available. Our result shows that cavity-assisted frequency conversion based on \mbox{PPKTP} is well suited for low-decoherence frequency conversion of quantum states of light.  
\end{abstract}

] 

\noindent

In quantum information science, low decoherence interfaces are required in order to map quantum information from one system
 to another. Light at near-infrared frequencies, i.e. around 1550\,nm, is well suited to distribute information over
 standard telecommunication fibre networks with low optical loss \cite{LNo08,Momo2010}. For quantum storage devices atomic
 transitions at around 800\,nm are actively researched, in the pulsed as well as in the continuous-wave 
regime  \cite{Julsgaard2004,HHGLBBL06,HSP10}. In order to link these frequency regimes, 
light at telecommunication wavelengths might be up-converted, e.g by second harmonic generation 
\cite{SHG,PhysRevLett.48.478,HKu92}. In past years a lot of progress in high-efficiency frequency up-conversion has been
 achieved. In \cite{Parameswaran2002} a pump depletion of 99\,\% was reported for 50\,ns pulses. In \cite{Pan2006} 
a single-photon conversion efficiency of 96\,\% was observed. Here, however, the efficiency of the quantum state 
transfer was lower than this value due to a 25\,\% background (dark) count rate. For second-harmonic generation of
 \emph{continuous-wave} (cw) light \cite{Ou1992,Paschotta1994} the highest external conversion efficiency reported so far 
is 90\,\%  \cite{TobiMeier2010}. This value also reflects the efficiency of the quantum state transfer since loss due to 
imperfect matching between input mode and cavity mode of the frequency converter was included and background and dark noise
 of the detection scheme was negligible. 

In this Letter we experimentally investigate the efficiency of cavity-assisted second-harmonic generation of continuous-wave
 light at 1550\,nm in \mbox{PPKTP}. This material is a promising candidate for reaching high quantum state transfer
 efficiencies, since the optical absorption of the material is low enough to observe up to 12.7\,dB of quantum noise
 squeezing \cite{Eberle2010}. We are in particular interested in maximizing the \emph{external} conversion efficiency,
 {as given by the power ratio of the 775\,nm cavity output beam and the 1550\,nm input beam}, i.e. our conversion
 efficiency value is not artificially increased by inferring to a situation with perfect mode matching into the cavity.
In the case of intra-cavity frequency doubling inside a laser resonator, a value of nearly 100\,\% was recently
 observed \cite{Zondy2010}.
 To the best of our knowledge, we report the highest external cw second-harmonic conversion efficiency observed so far. 
Our result is in full agreement with absolute values derived from a numerical model of the nonlinear cavity. 

Theoretically, frequency conversion of up to 100\,\% efficiency is possible in (lossless) {nonlinear optical materials}.
 In practice, a limitation occurs due to finite values of the nonlinearity of the material or of the optical pump power
 available, but in particular due to optical loss. A lack of pump power or nonlinearity can partially be compensated by 
using an optical resonator,  
but materials with low optical absorption eventually are essential for reaching high conversion efficiencies. Absorption 
directly reduces the conversion efficiency, but also limits the maximum value of the cavity finesse and the resonant field 
enhancement possible. Absorption also leads to heating, thermal lensing and a deformation of the cavity mode limiting 
the conversion efficieny \cite{LeTargat}.
\begin{figure}[!htbp]
\centerline{\includegraphics[width=8.6cm]{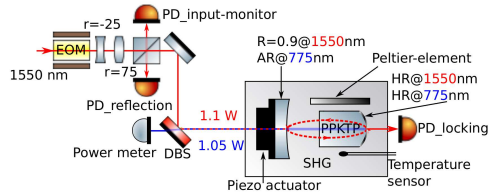}}
\caption{Schematic of experimental setup. SHG: second harmonic generation with nonlinear \mbox{PPKTP} crystal, DBS: dichroic
 beam splitter to separate 775\,nm and 1550\,nm light, PD\_input-monitor: photo detector monitoring power fluctuations of 
the 1550\,nm pump field, PD-reflection: photo detector for relative measurement of the conversion efficiency, PD\_locking:
 photo detector for the cavity length stabilization, EOM: electro-optic modulator producing sidebands of 138\,MHz for the 
cavity length stabilization.}
\label{exp-setup}
\end{figure}

In the experimental setup (figure \ref{exp-setup}) we started with continuous-wave light from a 1550\,nm fibre laser.
 After transmission through a three mirror ring resonator for spatial mode cleaning up to 1.1\,W were mode-matched into a 
standing-wave cavity containing a \mbox{PPKTP} crystal. A phase modulation was applied to the light field via an 
electro-optical modulator and used for active length stabilization of the cavity. Before the 1550\,nm laser entered the 
frequency doubling unit, an anti-reflective coated substrate reflected part of the beam onto photo diode 
\textit{PD\_input-monitor} in order to monitor {power drifts}. To separate the fundamental and harmonic fields, a dichroic
 beam splitter (DBS) with $\text{R}_{1550}>99.98\,\%$ and $\text{R}_{775}=0.7\,\%$ was also placed directly in front of the
 \mbox{PPKTP} cavity.
The \mbox{PPKTP} crystal \cite{Vyatkin2004} was plano-convex and had a dimension of $1\times2\times 9.3\,\text{mm}^{3}$ with a quasi phase-matching temperature of about $45\,^\circ\text{C}$.
 The curved surface had a radius of 12\,mm and was dielectrically coated yielding a reflectivity of about 99.95\,\% at 
the wavelengths of 1550\,nm and 775\,nm. The crystal's plane surface was anti-reflection coated (R $< 0.05\%$). We 
measured the absorption of the \mbox{PPKTP} to be $\alpha_{1550}<0.01\,\frac{\%}{\text{cm}}$ and
 $\alpha_{775}=(0.028\pm0.005)\,\frac{\%}{\text{cm}}$ for the fundamental and harmonic wavelengths, respectively, 
using the method described in \cite{Nico10}. The \mbox{PPKTP} cavity was completed by an external cavity mirror with 25\,mm radius of curvature and with reflectivities 
$\text{R}_{1550}=90.0\pm1.5\,\%$ and $\text{R}_{775}<0.2\,\%$. The mirror had a distance from the plane crystal surface of 
about 24\,mm, resulting in a cavity waist size of $\text{w}_{0}=37.6\,\mu\text{m}$
and was attached to a piezo electric transducer to scan and stabilize the resonator length. The measured mode matching of 
the TEM00 mode into the SHG cavity was better than 98\,\%, a value which sets an upper limit for our conversion efficiency.

\begin{figure}[!htbp]
\centerline{\includegraphics[width=9.6cm]{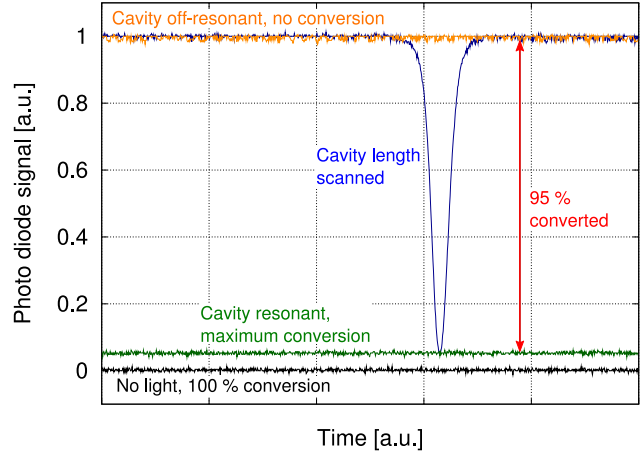}}
\vspace{-4mm}
\caption{Depletion measurement for a fundamental input power of 1.1\,W. Bottom trace (black): no light on \textit{PD\_reflection} refers to 100\,\% conversion. Top trace (orange): SHG cavity off-resonant refers to 0\,\% conversion. Trace at about 0.05 (green): SHG cavity is stabilized on maximum conversion. Peaked curve (blue): Scan of the cavity length over an Airy-peak of the TEM00-mode. The maximum conversion efficiency is $95\,\%$.}
\label{SHG-PD-Conversion}
\end{figure}
For low absorption and scattering losses, a pump depletion measurement is a feasible technique to determine the conversion
 efficiency \cite{Parameswaran2002}. In our setup, this was done by detecting the 1550\,nm light in reflection of the 
cavity using photo diode \textit{PD\_reflection}. For a cavity operated off-resonance the fundamental input field did not 
enter the cavity and hence gave a 0\,\% conversion efficiency reference on our photo diode \textit{PD\_reflection}. When 
the nonlinear cavity length was set on resonance, the fundamental field entered the SHG and some fraction of it was 
converted to its harmonic, which appeared as loss on the detected signal. 
A reference for maximum conversion (100\,\%) was measured, by blocking all light in front of the photo diode.  
Figure \ref{SHG-PD-Conversion} shows the result of our depletion measurement for a fundamental input power of about 1.1\,W.
 Apart from the two reference levels, the graph shows the depletion for a swept cavity length and for a cavity length 
controlled on maximum conversion. Note that for an increasing conversion the high over-coupling of the cavity got slightly
 reduced. We could quantify this effect by our simulation and also by a measurement of the less pronounced reduction of
 the faint light transmitted through the cavity. In order to take into account this effect we reduced our conversion 
efficiency value by 0.2 percentage points, however, this small correction was within the error bars of our final value
 of $(95\pm1)$\,\%.  The error bar was given by measurement statistics as well as the non-perfect anti-reflection coating 
of the crystal, uncertainties on the absorption and scattering losses inside the crystal, and related changes of the 
over-coupling. The photo detector nonlinearity was estimated to be $\sim0.1\,\%$ over the dynamic range used.

\begin{figure}[htb]
\centerline{\includegraphics[width=9.7cm]{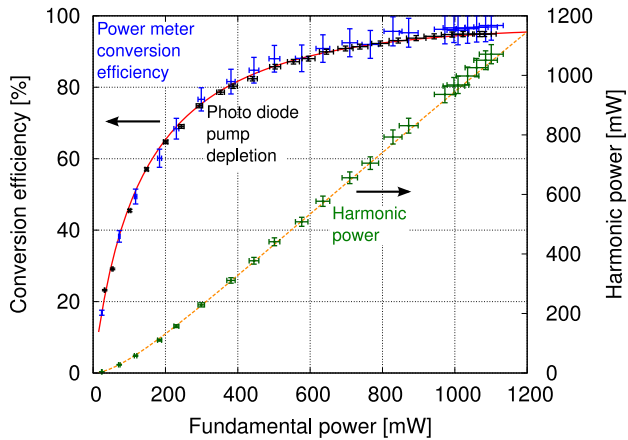}}
\vspace{-4mm}
\caption{Conversion efficiency and harmonic power (775\,nm) versus fundamental input power (1550\,nm). Data with small error bars (black): pump depletion measurements. Data with large error bars (blue and green): absolute power measurements at both wavelengths. The dotted orange line and the solid red line show the full simulation of our system based on independently determined parameters and a fitted crystal nonlinearity of 7.3\,pm/V. }
\label{Conversion}
\end{figure}
The data points with small error bars (black) in Fig.~\ref{Conversion} present depletion measurements for reduced input
 powers. The data perfectly fits to the result from our numerical time-domain simulation (solid red line) setting the 
effective nonlinearity of our crystal to a value of $d_{\rm{eff}}=7.3$\,pm/V, which is in good agreement with the slightly
 higher value at 1064\,nm as determined in \cite{PPKTP}. The simulation provides absolute power values for the second 
harmonic output field and thus for the conversion efficiency. It numerically applies a rigorous time-domain simulation of 
gaussian beams interacting with the non-linear medium inside an optical resonator. For this, the crystal is split into 
small segments of width $\Delta z$. Every segment's input and output modes (here fundamental and harmonic) are calculated
 analogous to a linear scattering mechanism. After each additional round trip the fields are compared with the preceding 
ones until a steady state is reached, which triggers the end of the simulation. For more details we refer to 
\cite{Nico-phd}. The input parameters of our simulation 
are the absolute input power at the fundamental wave length, the cavity mirror reflectivities, and cavity waist size, 
as well as the crystal specifications, such as length, absorption at the fundamental and harmonic frequency, and indices 
of refraction. 

In order to make an experimental consistency check we performed a second measurement using a commercial power meter 
(\emph{Ophir Optronics}) with a thermal measuring head and an absolute error bar of 3\,\% at both wavelengths. The data 
points are also shown in Fig.~\ref{Conversion} (in blue). This data is indeed consistent, however, has considerably larger
 error bars and was not used to determine the conversion efficiency value reported here. 

In {conclusion}, we observe an optical pump depletion of $(95 \pm 1)\%$ of 1.1\,W continuous-wave laser light at 1550\,nm 
through second harmonic generation in \mbox{PPKTP}. Based on independent measurements we estimate absorption and other 
optical losses to be insignificant with respect to our error bar. Due to energy conservation, the depletion value therefore
 corresponds to the external conversion efficiency. The result is consistent with independent but less accurate
 measurements with a calibrated power meter. We also find very good agreement with a numerical time-domain simulation of 
our experiment. The simulation shows that our conversion efficiency is limited by the non-perfect mode-matching and by the
non-perfect cavity impedance matching for the maximum input power available. At a pump power of 1.3\,W a maximum conversion of 
about 98\,\% should be reached.
From our results we expect that also nonclassical cw states of light can be frequency converted in \mbox{PPKTP} with similar efficiency. 

We acknowledge support from the International Max Planck Research School for Gravitational Wave Astronomy.


\begin{thebibliography}{10}
\newcommand{\enquote}[1]{``#1''}

\bibitem{LNo08}
M.-J. Li and D.~A. Nolan, J. Lightwave Technol. \textbf{26}, 1079 (2008).

\bibitem{Momo2010}
M.~Mehmet, T.~Eberle, S.~Steinlechner, H.~Vahlbruch, and R.~Schnabel, Opt.
  Lett. \textbf{35}, 1665 (2010).

\bibitem{Julsgaard2004}
B.~Julsgaard, J.~Sherson, J.~I. Cirac, J.~Fiurasek, and E.~S. Polzik, Nature
  \textbf{432}, 482 (2004).

\bibitem{HHGLBBL06}
M.~T.~L. Hsu, G.~H\'etet, O.~Gl\"ockl, J.~J. Longdell, B.~C. Buchler, H.-A.
  Bachor, and P.~K. Lam, Phys. Rev. Lett. \textbf{97}, 183601 (2006).



\bibitem{HSP10}
K.~Hammerer, A.~S. S\o{}rensen, and E.~S. Polzik, Rev. Mod. Phys. \textbf{82},
  1041 (2010).

\bibitem{SHG}
P.~A. Franken, A.~E. Hill, C.~W. Peters, and G.~Weinreich, Phys. Rev. Lett.
  \textbf{7}, 118 (1961).

\bibitem{PhysRevLett.48.478}
T.~F. Heinz, C.~K. Chen, D.~Ricard, and Y.~R. Shen, Phys. Rev. Lett.
  \textbf{48}, 478 (1982).

\bibitem{HKu92}
J.~Huang and P.~Kumar, Phys. Rev. Lett. \textbf{68}, 2153 (1992).

\bibitem{Parameswaran2002}
K.~R. Parameswaran, J.~R. Kurz, R.~V. Roussev, and M.~M. Fejer, Opt. Lett.
  \textbf{27}, 43 (2002).

\bibitem{Pan2006}
H.~Pan, H.~Dong, H.~Zeng, and W.~Lu, Appl. Phys. Lett. \textbf{89}, 191108
  (2006).


\bibitem{Ou1992}
Z.~Y. Ou, S.~F. Pereira, E.~S. Polzik, and H.~J. Kimble, Opt. Lett.
  \textbf{17}, 640 (1992).

\bibitem{Paschotta1994}
R.~Paschotta, P.~K\"{u}rz, R.~Henking, S.~Schiller, and J.~Mlynek, Opt. Lett.
  \textbf{19}, 1325 (1994).

\bibitem{TobiMeier2010}
T.~Meier, B.~Willke, and K.~Danzmann, Optics Letters \textbf{35}, 3742 (2010).

\bibitem{Eberle2010}
T.~Eberle, S.~Steinlechner, J.~Bauchrowitz, V.~H\"{a}ndchen, H.~Vahlbruch,
  M.~Mehmet, H.~M\"{u}ller-Ebhardt, and R.~Schnabel, Phys. Rev. Lett.
  \textbf{104}, 251102 (2010).

\bibitem{Zondy2010}
J.-J.~Zondy, F.~A.~Camargo, T.~Zanon, V.~Petrov and N.~U.~Wetter, Opt. Express \textbf{18}, 4796 (2010).

\bibitem{LeTargat}
R.~Le Targat, J.-J.~Zondy and P.~Lemonde, Opt. Commun. \textbf{247}, 471 (2005).

\bibitem{Vyatkin2004}
M.~Y. Vyatkin, A.~V. Avtlokhin, A.~G. Dronov, A.~A. Krylov, R.~L. Yagodkin,
  S.~V. Popov, J.~R. Taylor, and V.~P. Gapontsev, in \emph{Lasers and
  Electro-Optics} (2004).

\bibitem{Nico10}
N.~Lastzka, J.~Steinlechner, S.~Steinlechner, and R.~Schnabel, Appl. Opt.
  \textbf{49}, 5391 (2010).

\bibitem{PPKTP}
K.~Ehret, M.~Frede, S.~Ghazaryan, M.~Hildebrandt, E.-A. Knabbe, D.~Kracht,
  A.~Lindner, J.~List, T.~Meier, N.~Meyer, D.~Notz, J.~Redondo, A.~Ringwald,
  G.~Wiedemann, and B.~Willke, Nuclear Instruments and Methods in Physics
  Research Section A: Accelerators, Spectrometers, Detectors and Associated
  Equipment \textbf{612}, 83  (2009).

\bibitem{Nico-phd}
N.~Lastzka, \enquote{Numerical modelling of classical and quantum effects in
  non-linear optical systems}, Ph.D. thesis, Gottfried Wilhelm Leibniz
  Universit\"{a}t Hannover (2010).

\end{thebibliography}
\end{document}